\def\real{{\mathchoice
 {\hbox{$\displaystyle       \kern-.2mm 1\kern-1mm \mbox{\rm R}\kern-.2mm$}}
 {\hbox{$\textstyle          \kern-.2mm 1\kern-1mm \mbox{\rm R}\kern-.2mm$}}
 {\hbox{$\scriptstyle        \kern-.2mm 1\kern-1mm \mbox{\rm R}\kern-.2mm$}}
 {\hbox{$\scriptscriptstyle  \kern-.2mm 1\kern-1mm \mbox{\rm R}\kern-.2mm$}}}}

\def\complex{{\mathchoice
    {\hbox{$\displaystyle\kern-.2mm {\rm C}\kern-1.5mm\raise.2mm
                   \hbox{\vrule height6pt}\kern1.3mm$}}
    {\hbox{$\textstyle\kern-.2mm {\rm C}\kern-1.5mm\raise.3mm
                   \hbox{\vrule height6pt}\kern1.3mm$}}
    {\hbox{$\scriptstyle\kern-.2mm{\rm C}\kern-1.5mm\raise.2mm
                   \hbox{\vrule height3pt}\kern1.3mm$}}
    {\hbox{$\scriptscriptstyle\kern-.2mm{\rm C}\kern-1.5mm\raise.2mm
                  \hbox{\vrule height2pt}\kern1.3mm$}}}}

\documentclass{article}
\usepackage{epsfig}

\begin{document}
\setcounter{page}{1}
\title{Orbit Spaces of Compact Linear Groups}
\author{Vittorino Talamini}\date{INFN, Sezione di Padova, I-35131 Padova, Italy, and\\
Dip. di Ingegneria Civile, Universit\`a di Udine, I-33100 Udine, Italy\\
(e-mail: talamini@pd.infn.it)}

\markboth{V. Talamini}{Orbit Spaces of
 Compact Linear Groups}\maketitle

\begin{abstract}
The $\widehat P$-matrix approach for the determination of the
orbit spaces of compact linear groups enabled to determine all
orbit spaces of compact coregular linear groups with up to 4 basic
polynomial invariants and, more recently, all orbit spaces of
compact non-coregular linear groups with up to 3 basic invariants.
This approach does not involve the knowledge of the group
structure of the single groups but it is very general, so after
the determination of the orbit spaces one has to determine the
corresponding groups. In this article it is reviewed the main
ideas underlying the $\widehat P$-matrix approach and it is
reported the list of linear irreducible finite groups and of
linear compact simple Lie groups, with up to 4 basic invariants,
together with their orbit spaces. Some general properties of orbit
spaces of coregular groups are also discussed. This article will
deal only with the mathematical aspect, however one must keep in
mind that the stratification of the orbit spaces represents the
possible schemes of symmetry breaking and that the phase
transitions appear when the minimum of an invariant potential
function shifts from one stratum to another, so the exact
knowledge of the orbit spaces and their stratifications might be
useful to single out some yet hidden properties of phase
transitions. \footnote{ Published in the Proceedings of the 4th
International School on Theoretical Physics (SSPCM 96), Zajaczkowo
(PL), 29/8-4/9/1996, Eds. T. Lulek, W. Florek, B. Lulek, World
Sci. Singapore (1997), 204-222.}
\end{abstract}

\section{Basic Definitions on Compact Group Actions}

In this article it will be given only a short survey on the determination of
orbit spaces of compact linear groups. More details can be found in
\cite{6st,sar,stv}.

When a physical system has to show a symmetry of the nature, it must
be described mathematically on a certain representation space of a
group $G$ in such a way that all physically relevant
quantities are invariant with respect to $G$-transformations.\\
Often the group $G$ is compact and its representation space is finite
dimensional.
In this case in all generality one might suppose that $G$
is a group of real orthogonal matrices acting on
the real vector space $\real^n$.
In the following it will be assumed that $G \subseteq O(n)$ and also that
the origin of $\real^n$ is the only point left fixed by all transformations
of $G$.

The orbit through $x \in \real^n$ is the subset of $\real^n$ formed by all
points connected to $x$ by $G$-transformations:
$$\Omega(x)=\{g\cdot x,\quad \forall g \in G \},\qquad x \in \real^n$$
The isotropy subgroup $G_x$ of the point $x$ is the subgroup of $G$
that leaves $x$ fixed:
$$G_x=\{g \in G \mid g \cdot x = x \},\qquad x \in \real^n$$
All the points in a same orbit $\Omega(x)$ have isotropy subgroups in a same
conjugacy class $[G_x]$ of subgroups of $G$,
called the {\em orbit type} of $\Omega(x)$, in fact one has:
$$G_{g\cdot x}=g \cdot G_x \cdot g^{-1},\qquad \forall g\in G,\qquad x\in \real^n$$
To each orbit type $[H]$ it is associated
a {\em stratum} $\Sigma_{[H]}$, formed by all points of $\real^n$ that have
isotropy subgroups in $[H]$:
$$\Sigma_{[H]}=\{x\in \real^n \mid G_x \in [H]\}$$
The orbits (and the strata) are disjoint subsets of $\real^n$, as each orbit
has one and only one orbit type.\\
The orbits and the strata can be partially ordered according to their
orbit types [H]. The orbit type $[H]$ is said to be
{\em smaller} than the orbit type $[K]$: $[H]<[K]$,
if $H' \subset K'$ for some $H'\in [H]$ and $K'\in [K]$. Then
$[K]$ is {\em greater} than $[H]$.

Due to the compactness of $G$, the number of different orbit types is finite
and there is a unique minimal orbit type. The unique stratum of smallest
orbit type is called the {\em principal stratum}.
All other strata are called {\em singular}.

The {\em orbit space} is the quotient space $\real^n/G$ defined through the
equivalence relation relating points belonging to the same orbit.
The natural projection $\pi:\real^n \to \real^n/G$ maps the orbits of $\real^n$
into single points of $\real^n/G$. Projections of strata of $\real^n$
define strata of
$\real^n/G$. The principal stratum of $\real^n/G$ is always {\em open
connected} and {\em dense} in $\real^n/G$, so also $\real^n/G$ is connected.
If $[K]>[H]$, then $\pi(\Sigma_{[K]})$ lies in the boundary
of $\pi(\Sigma_{[H]})$,
so to greater orbit types correspond strata of smaller dimension and
the boundary of the principal stratum contains all singular strata.
Clearly there is a one to one correspondence between the strata of $\real^n$
and the strata of $\real^n/G$, so $\real^n$ and $\real^n/G$ are stratified
in exactly the same manner.

For all the $G$-invariant functions, $f(g\cdot x)=f(x),\ \forall g\in G,\ x\in
\real^n$, so the invariant functions are constant on the orbits. Then can then
be thought as functions defined on the orbit space and in this way one eliminates the
degeneration of all the points belonging to a same orbit, in which $f(x)$ is
constant.\\
The isotropy subgroup of the minimum point of an invariant potential function
determines the true symmetry group of a physical system and if this minimum
is not at the origin then one has a symmetry breaking.

The potential may depend on some variable parameters, so the location of
the minimum point also depends on these parameters and a
phase transition is realized when the minimum point changes stratum.
Keeping these things in mind one may realize that many properties of
invariant functions and phase transitions can be better
studied in the orbit spaces.

\section{Orbit Spaces in $\real^q$ and $\widehat P$-matrices}

A concrete mathematical description of the orbit space is achieved through
an {\em integrity basis} (IB) $\ p_1(x),\ldots,p_q(x)\ $ for the ring
$\ \real[\real^n]^G\ $ of the $G$-invariant
polynomial functions defined on $\real^n$. All $G$-invariant polynomial
(or $C^\infty$) functions can be expressed as polynomials (or $C^\infty$)
functions of the finite number $q$ of basic polynomial invariant functions
forming the IB:
$$f(x)=\widehat f(p_1(x),\ldots,p_q(x)),\qquad \forall f \in \real[\real^n]^G \qquad
(\mbox{or}\ \forall f \in C^\infty[\real^n]^G)$$
The IB is supposed minimal, i.e. no subset of the IB is itself an IB, and
formed by homogeneous polynomial functions.
The choice of the IB is not unique, but the group fixes the number $q$
of its elements and their degrees $d_1,\ldots,d_q$.\\
We assume that the basic invariants $p_i(x)$ are labelled in such a way that
$d_i \ge d_{i+1}$. As there are no fixed
points except the origin of $\real^n$, $d_q\geq 2$,
and, because of the orthogonality of $G$, we may take
$\ p_q(x)=\sum_{i=1}^n x_i^2\ $.

All IB transformations (IBTs):
$$p_i'(x) =p_i'(p_1(x),\ldots,p_q(x)),\qquad i=1,\ldots,q-1,\qquad p_q'(x)=p_q(x)$$
that satisfy the conventions adopted must have Jacobian matrix with elements
$J_{ij}(x)=\partial p_i'(x)/\partial p_j(x)$
that are 0 or $G$-invariant homogeneous polynomial functions of degree:
$\deg(J_{ij}) =d_i - d_j $.
Then, $J(x)$ is an upper block triangular matrix and $\det J(x)$
is a non vanishing constant.

The IB can be used to represent the orbits of $\real^n$ as points of
$\real^q$. In fact, given an orbit $\Omega$, the vector function
$(p_1(x), p_2(x),\ldots, p_q(x))\ $is constant on $\Omega$,
because the $p_i(x)$ are $G$-invariant. The $q$
numbers $p_i=p_i(x),\ x \in \Omega$, determine a point
$p=(p_1,p_2,\ldots,p_q)\in\real^q$, which can be
considered the image in $\real^q$ of $\Omega$.
No other orbit of $\real^n$ is represented in $\real^q$ by the same point
because the IB separates the orbits.\\
The vector map:
$$p:\real^n \to \real^q:x \to (p_1(x), p_2(x),\ldots,p_q(x))$$
is called the {\it orbit map}. It maps $\real^n$ onto the subset
${\cal S}\subset \real^q$:
$${\cal S}=p(\real^n) \subset\real^q$$
such that each orbit of $\real^n$ is mapped in one and only one point of ${\cal S}$.\\
The orbit map $p$ induces a one to one correspondence
between $\real^n/G$ and ${\cal S}$ so that ${\cal S}$ can be concretely
identified with the orbit space of the $G$-action.\\
${\cal S}$ is a closed connected semi-algebraic
proper subset of $\real^q$ stratified in exactly the same manner as $\real^n$.
All the strata $\sigma$ of ${\cal S}$ are images of the strata $\Sigma$
of $\real^n$ through the orbit map and if
$\Sigma'$  is of greater orbit type than $\Sigma$,
then $\sigma'=p(\Sigma')$ lie in the boundary of $\sigma=p(\Sigma)$.
The interior of ${\cal S}$ hosts the principal stratum and all singular strata
lie in the bordering surface of ${\cal S}$. Like all semi-algebraic sets ${\cal S}$ is
stratified in primary strata and each primary stratum is the image of a
connected component of a stratum of $G$.

The origin of $\real^n$ is the only stratum of the greatest orbit
type $[G]$ and its image through the orbit map is always the origin of
$\real^q$, because of the homogeneity of the IB. The origin of
$\real^q$ lies then in the boundary of all other strata of ${\cal S}$.\\
${\cal S}$ is unlimited because $\forall x\in \real^n$ the points $x$ and
$\lambda x$, $\forall \lambda\in \real,\ \lambda\neq 0,\ $ belong to the same
stratum because of the linearity of the $G$-action, so,
as $x$ belongs to the sphere with equation $p_q(x)=(x,x)=x^2$,
all the positive $p_q$ axis of $\real^q$ must belong to ${\cal S}$.
Then any spheric surface of $\real^n$ with equation $(x,x)=r^2>0$ intersects
all strata of $\real^n$ except the origin.
Then any plane $\Pi_r$ of $\real^q$ with equation $p_q=r^2>0$ intersects all
strata of $\real^q$ except the origin. As the sphere is a compact set,
${\cal S}\cap \Pi_r$, gives a compact connected section of the orbit space ${\cal S}$.
This section is sufficient to imagine the whole shape of ${\cal S}$, because
going down to the origin in the direction of the $p_q$ axis this section must
contract to reduce at the end to the origin point and
going up to infinite this section must expand, mantaining in any case its
topological shape.

As an example, Figure 1 shows the orbit space classified in
\cite{6st} as III.2 (Table~\ref{tabR} lists the coregular groups
that have this orbit space) and a section of this orbit space with
a plane $p_3=\mbox{constant}$.
\begin{figure}[h]
\vfill
   \begin{minipage}{0.5\linewidth}
\hskip 0.7cm
\includegraphics{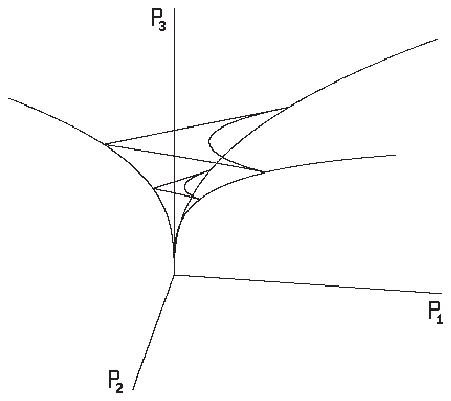}
   \end{minipage} \hfill
   \begin{minipage}{0.5\linewidth}
   \vskip 0.2cm
\includegraphics{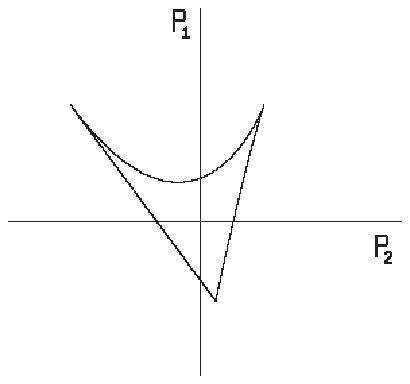}
   \end{minipage} \hfill
\caption{\protect\footnotesize Orbit space III.2 and its section
 with a plane $p_3=\mbox{constant}$.}
\end{figure}

All $G$-invariant $C^\infty$ functions $f(x)$, defined on
$\real^n$, can be expressed as $C^\infty$ functions of the basic
invariants, so they define $C^\infty$ functions $\widehat f(p)$,
defined on $\real^q$:
$$f(x) = \widehat f(p_1(x),\ldots,p_q(x))  \to \widehat f(p_1,\ldots,p_q)$$
The functions $\widehat f(p)$ are defined also in
points $p\notin {\cal S}$ but only the restriction $\widehat f(p)\mid_{p\in {\cal S}}$
has the same range as $f(x),\  x \in \real^n$, in fact $f(x)=f(p)$ if
$p=p(x)$, that is if $p\in {\cal S}$.\\
All $G$-invariant $C^\infty$ functions can then be studied in the orbit
space ${\cal S}$ but one needs to know
exactly all equations and inequalities defining ${\cal S}$ and its strata.\\
A polynomial $f(p)$ is said {\it $w$-homogeneous} of {\it weight} $d$
if the polynomial $f(p(x))$ is homogeneous with degree $d$.
Each coordinate $p_1,\ldots,p_q$ of $\real^q$ has
then a weight $d_1,\ldots,d_q$.

The IBTs can be viewed as coordinate transformations of $\real^q$:
$$p_i' =p_i'(p_1,\ldots,p_q),\qquad
           i=1,\ldots,q-1,\qquad p_q'=p_q$$
The Jacobian matrix $J(p)$ inherits the
properties of $J(x)$, in particular its matrix elements
$J_{ij}(p)=\partial p_i'(p)/\partial p_j$ are 0 or
$w$-homogeneous polynomials in $p$ of weight $d_i-d_j$ and $\det J(p)$ is
a non vanishing constant.
The only coordinate transformations of $\real^q$
of interest are those corresponding to IBTs and they are called
again IBTs (of $\real^q$).\\
The IBTs change the form of ${\cal S}$ but not its topological shape and
stratification.

An IB is said {\em regular} if it does not exist any polynomial function
$f(p)$ such that:
$$f(p_1(x),\ldots,p_q(x))=0\qquad \forall x \in \real^n$$
Otherwise the IB is said {\em non regular}. Obviously $f(p)$
cannot be solved with respect of any of the variables, otherwise the basis
would not be minimal.\\
A linear group $G$ with a regular $IB$ is said {\em coregular}.
Otherwise it is said {\em non-coregular}.\\
If the basis is non regular then generally it exists an ideal of
polynomial relations between the elements of the IB and the (independent)
generators of this ideal $f_1,\ldots,f_r$ can always be chosen homogeneous.
The orbit space ${\cal S}$ must then be contained in the surface ${\cal Z}$ of
$\real^q$ defined by the equations:
$$f_1(p)=0,\ldots,\quad f_r(p)=0$$
and has the same dimension $q-r$ of ${\cal Z}$. ${\cal Z}$ is called the
{\em surface of the relations} and the couple $(q,q-r)$ the
{\em regularity type} (or $r$-type) of the IB and of the group $G$.\\
When $G$ is coregular there are no relations, $r=0$, and ${\cal Z}\equiv \real^q$,
so ${\cal S}$ is $q$-dimensional.

In a point $x\in\Sigma \subset \real^n$, the number of
linear independent gradients of the basic invariants is equal to the
dimension of the stratum $p(\Sigma)\subset {\cal S}$.
It is convenient to construct the $q \times q$
Grammian matrix $P(x)$ with elements $P_{ab}(x)$
that are scalar products between the gradients of the basic invariants:
$$   P_{ab}(x) = (\nabla p_a(x),\nabla p_b(x))$$
$P(x)$ is then positive semidefinite $\forall x \in \real^n$, and for
$x\in \Sigma \subset \real^n$ $ \mbox{rank} P(x)$ equals the dimension of the stratum
$p(\Sigma)$ of ${\cal S}$.\\
Because of the covariance of the gradients of $G$-invariant
functions ($\nabla f(g \cdot x)=g \cdot \nabla f(x)$)
and of the orthogonality of $G$ (which implies the invariance of
the scalar products),
the matrix elements
$P_{ab}(x)$ are $G$-invariant homogeneous polynomial functions of
degree $d_a+d_b-2$.
Then, all the matrix elements of $P(x)$ can be expressed as
polynomials of the basic invariants.
One can then define a matrix $\widehat P (p)$ in
$\real^q$ such that:
$$    P_{ab}(x) =\widehat P_{ab}(p_1(x),\ldots,p_q(x))=\widehat P_{ab}(p)
\quad \forall x \in \real^n\ \mbox{and}\ p=p(x)$$
At the point $p=p(x)$, image in $\real^q$
of the point $x\in\real^n$ through the orbit map, the matrix $\widehat P(p)$
is the same as the matrix $P(x)$.
The matrix $\widehat P(p)$ is however defined in all $\real^q$, also
outside ${\cal S}$, but only in ${\cal S}$ it reproduces $P(x),\ \forall x\in\real^n$.\\
The properties of the matrix $\widehat P(p)$ depend on the definition of
$P(x)$ and are the following:
\begin{enumerate}
\item $\widehat P(p)$ is a real, symmetric $q \times q$ matrix.
\item  $\widehat P(p)$ is positive semidefinite in ${\cal S}$.
${\cal S}$ is the {\em only} region of ${\cal Z}$ where $\widehat P(p)\geq 0$.
\item $\mbox{rank}\widehat P(p)$ equals the dimension of the stratum
containing $p$.
\item the matrix elements
$\widehat P_{ab}(p)$ are $w$-homogeneous polynomial functions of weight
$d_a+d_b-2$ and the last row and column of $\widehat P(p)$ have the fixed form:
$$\widehat P_{qa}(p) =\widehat P_{aq}(p) = 2 d_a p_a\qquad \forall a=1,\ldots, q$$
\item $\widehat P(p)$ transforms as a contravariant tensor
under IBTs:
$$  P_{ab}'(p') = J_{ai}(p) J_{bj}(p) P_{ij}(p)$$
\end{enumerate}
The matrix $\widehat P(p)$ completely determines ${\cal S}$ and its stratification.
Defining ${\cal S}_k$ the union of all $k$-dimensional strata of ${\cal S}$, one has:
$${\cal S}=\{p\in {\cal Z} \mid \widehat P(p)\geq 0\}$$
$${\cal S}_k=\{p\in {\cal Z} \mid \widehat P(p)\geq 0,\ \mbox{rank} P(p)=k\}$$
The principal stratum has dimension $q-r$, equal to the maximum rank of
$\widehat P(p)$, with $r$ the number of independent relations among the $p_i(x)$,
and coincides with ${\cal S}_{q-r}$.
(Obviously if $G$ is coregular $r=0$ and ${\cal Z}\equiv \real^q$).\\
To find out ${\cal S}_k$ one may impose that all the principal minors of $\widehat P(p)$
of order greater than $k$ have zero determinant and that at least one of
of those of order $k$ have non zero determinant.
From this one sees that ${\cal S}$ is semialgebraic because it is defined through
algebraic equations and inequalities.\\
In order to classify orbit spaces, it is sufficient
to classify the corresponding matrices $\widehat P(p)$, as they determine ${\cal S}$
completely and this has been done in \cite{6st,7st}.

\section {Determination of the Allowable $\widehat P$-matrices of
Compact Linear Groups through the Canonical Equation}

The polynomials defining the singular strata of ${\cal S}$ (and also the principal
stratum if $G$ is non-coregular) must satisfy a differential relation
(that characterize them) that has been crucial in \cite{6st,7st,stv}
to determine the $\widehat P$-matrices without the explicit use of
the IB's. Let's see the origin of this relation.\\
Let $\sigma$ be a primary stratum of ${\cal S}$ and $I(\sigma)$ the ideal of the
polynomials defined in $\real^q$ vanishing in $\sigma$.
Then $\Sigma=\{x\in \real^n \mid p(x)\in \sigma \}$ is a connected component
of a stratum of $\real^n$. For all $\widehat f\in I(\sigma)$,
$f(x)=\widehat f(p(x))$ is a $G$-invariant function that vanishes
on $\Sigma$.
Then in all regular points $x$ of $\Sigma$ one must have that
$\ \nabla f(x) |_{x\in \Sigma}\; \perp \; \Sigma\ $
because $f(x)$ is constant in $\Sigma$, and that
$\ \nabla f(x) |_{x\in \Sigma}\ $ is tangent to $\Sigma$
because $f(x)$ is $G$-invariant and the gradients of $G$-invariant functions
are tangent to the strata. Then the only possibility is that:
$$\nabla f(x) |_{x\in \Sigma}\ = \ 0$$
Applying the partial differentiation rule and taking the scalar product
with $\nabla p_a(x)$, $a=1,\ldots,q$, one obtains $q$ relations expressed
only in terms of the $p_i(x)$, that can so be defined also in $\real^q$:
$$0=\nabla f(x) |_{x\in \Sigma}=
\sum_{b=1}^q \nabla p_b(x)
\left. \frac{\partial f(x)}{\partial p_b(x)}\right|_{x\in \Sigma}=
\sum_{b=1}^q (\nabla p_a(x),\nabla p_b(x))
\left.\frac{\partial f(x)}{\partial p_b(x)}\right|_{x\in \Sigma}=$$
$$=\sum_{b=1}^q {\widehat P}_{ab}(p(x))
\left.\frac{\partial f(p(x))}{\partial p_b(x)}\right|_{x\in \Sigma}=
\sum_{b=1}^q {\widehat P}_{ab}(p)
\left.\frac{\partial f(p)}{\partial p_b}\right|_{p\in \sigma},\qquad a=1,\ldots,q$$
This means that:
$$\left.\sum_{b=1}^q {\widehat P}_{ab}(p)
\partial_b f(p)\right|_{p\in \sigma}\in I(\sigma),\qquad a=1,\ldots,q$$
where $\partial_b$ means partial derivation with respect to $p_b$.\\
It is convenient to distinguish two cases:
\begin{enumerate}
\item
$I(\sigma)$ has only one generator $a(p)$.
In this case one obtains the following relations:
$$\sum_{b=1}^q {\widehat P}_{ab}(p) \partial_b a(p) =
\lambda_a(p) a(p) \qquad a=1,\ldots,q$$
with the $\lambda_a(p)$ $w$-homogeneous polynomials of weight $d_a-2$.
In this case $\sigma$ is a surface of dimension $q-1$ and its intersection
with the region $\widehat P(p)\geq 0$ gives a
singular strata of maximal dimension if $G$ is coregular or the
principal stratum if $G$ is non-coregular of $r$-type $(q,q-1)$.
In this last case the equation $a(p)=0$ defines ${\cal Z}$.
\item
$I(\sigma)$ has more independent generators $a^{(1)}(p),\ldots, a^{(s)}(p)$.
In this case one obtains the following relations:
$$\sum_{b=1}^q {\widehat P}_{ab}(p) \partial_b a^{(j)}(p)=
\sum_{i=1}^s \lambda_{a}^{(i,j)}(p) a^{(i)}(p) \qquad a=1,\ldots,q,\ j=1,\ldots,s$$
with the $\lambda_{a}^{(i,j)}(p)$ $w$-homogeneous polynomials of weight
$d_a-2+w(a^{(j)})-w(a^{(i)})$.
$\sigma$ in this case is a surface of dimension $q-s$.
It can be a principal stratum only if $G$ is non-coregular of $r$-type
$(q,q-s)$. In all other cases it is a singular stratum.
\end{enumerate}
Both the two relations written are called {\em master relations}.
At the moment only the case of $(q-1)$-dimensional strata has been
investigated: the coregular case is presented in \cite{6st} and
the non-coregular case of $r$-type $(q,q-1)$ in \cite{stv} and all what
follows concerns only $(q-1)$-dimensional strata.

It has been proved in \cite{6st} that all irreducible polynomials $a(p)$
that satisfy the master relation must be factors of $\det{\widehat P}(p)$.
Any product of them satisfies the master relation too.
All these polynomials are called {\em active}.
The product $A(p)$ of all irreducible active polynomials is called
the {\em complete} (active) factor of $\det{\widehat P}(p)$.
The surface $\sigma=\{p\in {\cal Z} \mid \widehat P(p)\geq 0, A(p)=0\}$ coincides
with the whole boundary of ${\cal S}$ if $G$ is coregular
(${\cal Z}\equiv \real^n$ in this case) or coincides with
the whole principal stratum of ${\cal S}$ if $G$ is non-coregular of $r$-type $(q,q-1)$.
$\det {\widehat P}(p)$ may contain a non active factor $B(p)$ that is called
{\it passive}. $A(p)$ and $B(p)$ are uniquely defined except by
non-zero constant factors.\\
In \cite{6st} are studied the properties of the
master relation with respect to IBTs, and the main results are the following:
\begin{enumerate}
\item In some IB, called {\em $A$-bases}, the master relation
has the following {\it canonical} form:
$$\sum_{b=1}^q \widehat P_{ab}(p) \partial_b A(p) = 2 w(A)\delta_{aq}A(p) \qquad a=1,\ldots,q$$
The case $a=q$ is just a homogeneity condition on $A(p)$ and only the cases
$a=1,\ldots,q-1$ are characteristic of the $(q-1)$-dimensional strata.
The variable $p_q$ can then be easily eliminated from the equation if one
restricts to a plane $\Pi$ of constant $p_q$, for example
$\Pi=\{p \in \real^q \mid p_q=1\}$.
\item
In all $A$-bases, the restriction $A(p)\mid_{p\in\Pi}$,
has at most one local non-degenerate extremum in $\Pi$, outside of the region
$A(p)\mid_{p\in\Pi}\;=0$, at the point $p_{0}=(0,\ldots,0,1)$.
\end{enumerate}
If $G$ is coregular one has also the following results:
\begin{enumerate}
\item[3.]
$p_{0}$ always exists and lies in the interior of ${\cal S}$.
\item[4.]
In some $A$-bases, the {\it standard} $A$-bases, $\widehat P(p)$
evaluated at $p_0$ is diagonal.
\end{enumerate}
Properties 3. and 4. exclude linear terms in
$A(p)\mid_{p\in\Pi}$ and requires that all the quadratic terms $p_i^2$
have coefficients of equal sign (negative if one requires $A(p)\mid_{p\in\Pi}$
to be maximum at $p_0$) and that there are no mixed quadratic terms $p_i p_j$.
The weight $w(A)$ of $A(p)$ is then bounded:
$2 d_1 \leq w(A)\leq w(\det \widehat P) = 2 \sum_{i=1}^q (d_i-1)$.\\
If $G$ is non-coregular properties 3. and 4. are not valid but $A(p)$ must
satisfy in addition a second order boundary condition \cite{stv}.

Given the IB, and following the lines reported above,
it is easy to determine the matrix $\widehat P(p)$, the complete factor $A(p)$,
and the subset ${\cal S}$ of $\real^q$ that represents the orbit space of the
group action. The $\widehat P$-matrices can however be calculated and classified
without the knowledge of the IB's.
The many conditions found on the form of
a general $\widehat P$-matrix and on the complete factor
$A$ allow to find out all possible solutions to the
{\em canonical equation} that are compatible with these
conditions (the master relation in its canonical form is
used now to find out the $\widehat P$-matrices, so it is better to call it equation
instead of relation).

In \cite{6st,7st} the canonical equation is solved for the case of coregular
groups. We only fixed the number $q$ of the basic invariants and we
considered all matrix elements
$\widehat P_{ab}(p)$ and $A(p)$ and $B(p)$ as unknown $w$-homogeneous polynomials
satisfying the following {\em initial conditions}:
\begin{enumerate}
\item $\widehat P(p)$ is a real, symmetric $q \times q$ matrix;
\item the matrix elements
$\widehat P_{ab}(p)$ are $w$-homogeneous polynomial functions of weight
$d_a+d_b-2$ and the last row and column of $\widehat P(p)$ have the fixed form:
$$\widehat P_{qa}(p) =\widehat P_{aq}(p) = 2 d_a p_a\qquad \forall a=1,\ldots, q$$
\item  $\widehat P(p_0)$ is diagonal and positive definite;
\item $A(p)$ is $w$-homogeneous of weight $w(A)$ such that:
$2 d_1\leq w(A)\leq w(\det{\widehat P})$.
Its restriction to the plane $p_q=1$ has no linear terms and
only quadratic terms of the type $-k_i p_i^2$ with $k_i>0$.
\item $A(p)$ is a factor of $\det{\widehat P}$, so the equation
$A(p)B(p) = \det{\widehat P}(p)$ defines $B(p)$.
\item the canonical equation must be satisfied.
\end{enumerate}
In all these unknown polynomials the dependence
on the variable $p_1$ can always be rendered explicit even if all
degrees $d_i,\ i\neq q$ are unknown.
Items 5. and 6. give a system of coupled differential equations that
should be solved by unknown $w$-homogeneous polynomial functions.
The initial conditions imposed
were so strong that this system could be solved analytically and it
gave only a finite number of different solutions for each value
of the dimension $q$ ($q$ = 2, 3, 4, but there is no reason to believe that
this will not be true for higher values of $q$).
The matrices $\widehat P (p)$ that together with the corresponding complete factor
$A(p)$ satisfy these initial conditions, are such that $\widehat P(p)\geq 0$
only in a connected subset ${\cal S}$ of $\real^q$ (whose boundary satisfy the
equation $A(p)=0$) and are called {\em allowable $\widehat P$-matrices}.
The name originates from the fact that they are potentially determined by
an IB of an existing group $G$, but it is not known in general if
that group does really exist. It is clear however that all $\widehat P$-
matrices determined by the IB of the existing compact coregular
groups are allowable $\widehat P$-matrices.

This approach has been recently applied with success to the determination of
the allowable $\widehat P$-matrices associated to compact non-coregular linear groups
of $r$-type $(q,q-1)$ \cite{stv}.
The initial conditions 3. and 4. for the case of coregular groups are no longer
valid and must be replaced by the following two conditions:
\begin{enumerate}
\item [3.] $A(p)$ is $w$-homogeneous of weight $w(A)\leq w(\det \widehat P)$.
\item [4.] $A(p)$ satisfies the second order boundary condition \cite{stv}.
\end{enumerate}
Then one proceeds as in the coregular case. Here not all solutions found
are such that $\widehat P(p)\geq 0$ in a connected subset of $\real^q$, so, a
posteriori, one has to discard some solutions.
It can however be proved that when the point
$p_0$, where $A(p)\mid_{p\in\Pi}$ has an extremum, exists and does not
belong to ${\cal Z}$, then the allowable $\widehat P$-matrices of the
non-coregular groups of $r$-type $(q,q-1)$ are necessarily allowable
$\widehat P$-matrices of coregular groups with $q$ basic invariants.
In this case the orbit spaces ${\cal S}'$ of the non-coregular groups lie always
in some connected $(q-1)$-dimensional singular stratum of the orbit space
${\cal S}$ of a coregular group.

In table 1 I report all 2-, 3- and 4-dimensional
allowable $\widehat P$-matrices for coregular groups,
showing the corresponding weights $[d_1,\ldots,d_{q-1},2]$
of the $p_i$ and the weight
$w(A)$ of the complete factor $A(p)$. The parameters $j_i$ and $s$ that
appear in the table are arbitrary positive integers limited only by
$d_1\geq \cdots \geq d_{q-1} \geq 2$.\\ The explicit forms of
these allowable $\widehat P$-matrices are given in \cite{6st,7st}.\\
These allowable $\widehat P$-matrices share the following
properties:
\begin{enumerate}
\item For each number $q$, there are a finite number of classes of
allowable $q\times q$ $\widehat P$-matrices.
In each class the $\widehat P$-matrices differ only in some positive integer
parameters $j_i$ and in a scale factor $s$. These parameters
fix also the values of the degrees $[d_1,d_2,\ldots,d_q-1]$. In
$\Pi=\{p\in \real^q | p_q=1\}$ all the matrices that differ only for the
value of $s$ become identical.
\item
In convenient $A$-bases all coefficients of the
$\widehat P$-matrix elements are integer numbers.
\end{enumerate}
In \cite{7st,t1} the classes $A9(j_1)$ and $A10(j_1)$ were forgotten.
These classes of solutions have been determined by applying the induction rules
\cite{st} to the case $q=3$. These induction rules permit to write down
easily most of the solutions for the $(q+1)$-dimensional coregular case
once one knows those of the $q$-dimensional case.
The 4 induction rules discovered up to now when applied to the solutions
corresponding to $q=2,3$ give all solutions of the case $q=3,4$, except
when the complete factor contain a term in $p_1^q$ (and this seems the case
of all irreducible finite groups generated by reflections) and except the
class $D2$ (that probably can be derived from the class III.3 with
an induction rule not yet discovered). The induction rules probably reflect
some properties of groups but they are not yet understood.

In \cite{stv} are determined all allowable $\widehat P$-matrices for non-coregular
groups of $r$-type $(q,q-1)$, tht is groups
with 3 invariants among which there is one algebraic relation. It results
only one class of allowable $\widehat P$-matrices for these groups
and these $\widehat P$-matrices are the same as those for the coregular case with
3 invariants classified in \cite{6st} as $I(1,1)$.
The only difference is that the principal stratum now is the 2-dimensional
surface that makes up the boundary of ${\cal S}$ in the coregular case.
It is possible that all orbit spaces of non-coregular groups of $r$-type
$(q,q-1)$ occour at the border of the orbit spaces of coregular groups.

\begin{table}[h]\begin{center}\label{tab0}\caption{ALLOWABLE  $\widehat P$-
MATRICES OF COREGULAR GROUPS OF DIMENSION $q = 2, 3, 4$.} \vskip
0.4cm\begin{tabular}{|c|l|c|c|}              \hline
$q$&CLASS&      $w(A)$        &           $[d_1,\ldots,d_q]$     \\
\hline
2&$I_2$   &     $ 2 d_1$       &           $[s,2]$                    \\
\hline
3&$I(j_1,j_2)$              &
$  2 d_1 $       &    $    [s(j_1+j_2)/2, s,2]$\\
&$II(j_1)  $               &   $ 2 d_1+d_2 $    &     $
[s(j_1+1), 2s,2]$                  \\
&$III.1$    &    $ 3
d_1   $     &   $        [4s, 3s,2]$                     \\
&$III.2$
   & $   3 d_1   $      &   $        [6s, 4s,2]$
\\
&$III.3$    & $     3 d_1  $     &   $       [10s,6s,2]$                     \\
\hline
4&$A1(j_1,j_2,j_3,j_4)$     &
$  2 d_1   $    &$[s(j_1+j_2)(j_3+j_4)/4, s(j_1+j_2)/2,s,2]$\\
&$A3(j_1,j_2,j_3)$         &      $ 2 d_1+d_3  $& $
[s(j_1+1)(j_2+j_3)/2, s(j_1+1), 2s,2] $ \\
&$A2(j_1,j_2,j_3,j_4)$ &   $
2 d_1    $       & $  [s((j_1+1)(j_2+j_3)/2+j_4), s(j_1+1), 2s,2]$\\
&$A4(j_1,j_2)    $         & $ 2 d_1+j_1 d_3$   & $
2s j_2, s(j_1+j_2), 2s,2]     $   \\
&$A5(j_1,j_2,j_3)$         & $   2
d_1+d_2  $   & $ [s (j_1(j_2+1)+j_3), 2s(j_2+1), 2s,2]    $\\
&$A6(j_1,j_2,j_3)$         & $   2 d_1+d_2  $   & $
[s(j_1+j_2)(j_3+1)/2, s(j_1+j_2), s,2]$ \\
&$A7(j_1,j_2)$             & $
2 d_1+d_2+d_3$   & $        [s j_1(j_2+1), 2s j_1, 2s,2]   $\\
&$A8(j_1,j_2)$             & $  2 d_1+2 d_2 $   &$[s(j_1+1), s(j_2+1), 2s,2]     $      \\
&$A9(j_1)$             & $  2 d_1+ d_3 $   &$[2s(j_1+1), s(j_1+1), 2s,2]     $      \\
&$A10(j_1)$            & $  2 d_1+ d_2 $   &$[s(j_1+1), 2s, s,2]     $      \\
&$B1(j_1)    $             & $   2
d_1     $    & $         [6s j_1, 4s, 3s,2]           $     \\
&$B2$             & $  3 d_1      $    & $         [4s, 3s, 3s,2]
$     \\
&$B3(j_1,j_2)$             & $    3 d_1    $    & $
[2s(j_1+j_2), 3s(j_1+j_2)/2, s,2]$     \\
&$B4(j_1)    $             & $
3 d_1+d_3   $    & $        [4s j_1, 3s j_1, 2s,2]        $\\
&$C1(j_1,j_2)$             & $    2 d_1    $    & $
[3s(j_1+2j_2), 6s, 4s,2]      $      \\
&$C2(j_1)    $             & $ 2
d_1+d_2   $    & $         [6s j_1, 6s, 4s,2]          $      \\
&$C3(j_1)
$             & $ 2 d_1+2 d_2 $    & $       [3s(j_1+1), 6s, 4s,2]$      \\
&$C4         $             & $  3 d_1      $
& $        [6s, 4s, 3s,2]              $      \\
&$C5(j_1,j_2)$
& $    3 d_1    $    & $   [3s(j_1+j_2), 2s(j_1+j_2), s,2] $\\
&$C6(j_1)    $             & $ 3 d_1+d_3   $    & $        [6s j_1,4s j_1, 2s,2]        $      \\
&$D1(j_1)    $             & $    2 d_1
$    & $        [15s j_1, 10s , 6s,2]         $      \\
&$D2         $
& $   3 d_1     $    & $         [10s, 6s, 4s,2]            $
\\
&$D3(j_1,j_2)$             & $    3 d_1    $    & $
[5s(j_1+j_2), 3s(j_1+j_2), s,2] $      \\
&$D4(j_1) $                & $
3 d_1+d_3  $    & $         [10s j_1, 6s j_1, 2s,2]   $         \\
&$E1$
   &   $  4 d_1  $      &$            [5s, 4s, 3s,2]
$          \\
&$E2$       &   $ 4 d_1   $      &$
[6s, 4s, 4s,2]      $           \\
&$E3$     &   $ 4
d_1   $      &$            [8s, 6s, 4s,2]      $           \\
&$E4$
  &  $  4 d_1  $       &$           [12s, 8s, 6s,2]
$           \\
&$E5$      &  $  4 d_1  $       &$ [30s, 20s, 12s,2]     $            \\ \hline
\end{tabular}\end{center}
{\bf Notation}.
$q$: number of basic invariants and dimension of the $\widehat P$-matrices.
CLASS: class of $\widehat P$-matrices in the notation of \cite{6st,7st}.
$w(A)$ degree of the active factor $A(p)$ determining the boundary of the
orbit space.
$[d_1,\ldots,d_q]$: degree of the basic invariants.

\end{table}

\section{Orbit Spaces of Coregular Compact Groups}

All linear compact groups generate a $\widehat P$-matrix that must be an
allowable $\widehat P$-matrix. The converse may also be true but one must find
the linear groups corresponding to a given allowable $\widehat P$-matrix.

In the case of irreducible finite groups generated by reflections the IB
are well known and when these IB contain 2, 3 and 4 basic invariants
the corresponding $\widehat P$-matrices have been determined \cite{sv}.
All these $\widehat P$-matrices, after a proper IBT, exactly coincide with
those in the classification of allowable $\widehat P$-matrices reported
in \cite{6st,7st}.

In the case of compact Lie groups the IB is known only in few cases.
G. W. Schwarz \cite{sch} gave a classification of all complex coregular
representations of complex simple Lie groups,
together with the number and degrees of the basic invariants,
and for some groups also some hints to write down the integrity basis.\\
From this classification it is possible to deduce all real coregular
representations of compact simple Lie groups. Here under I shall sketch
how this has been possible. Proofs and details can be found in \cite{t0,stv2}.

\begin{itemize}
\item
Let $G$ be a compact Lie group and $G_c$ be its {\it complexification}.
$G_c$ is the {\em smallest} complex Lie group that
contains $G$ and $G$ is a {\it maximal compact subgroup} of $G_c$.
\item
Let $\varphi$ be a real representation of $G$ in a real vector space
$W_{\varphi }$. $\varphi$ defines uniquely a complex representation of $G_c$
in the complex space $V_{\varphi}=\complex \otimes W_{\varphi}$.
Viceversa, a complex representation $\varphi$ of $G_c$ in the complex space
$V_{\varphi }$ defines a complex representation of $G$ in $V_{\varphi }$.
All these representations are completely reducible and
$G$ and $G_c$ are then {\em reductive} groups.
\item
Every complex reductive Lie group may be identified with a complex linear
algebraic group so that its complex analytic representations coincide
exactly with its rational representations. Then the linear group
$\varphi(G_c)$ is a complex algebraic group as well as a complex Lie group.
\item
Every linear group $H\subset GL(n,\complex)$ and its
Zariski closure $\mbox{cl}(H)$ have the same polynomial invariants:
$$\complex[\complex^n]^H=\complex[\complex^n]^{\mbox{{\small cl}}(H)}$$
\item
The linear group $\varphi(G_c)$ is the Zariski closure of the linear group
$\varphi(G)$, so $\varphi(G)$ and $\varphi(G_c)$
have equal rings of invariant polynomials, and in particular these rings
have the same IB.
\item
Some complex representations $\varphi$ of $G_c$ admit a {\em real form} for the
maximal compact subgroup $G$ and in this case the linear group $\varphi(G)$
is equivalent to a group of real orthogonal matrices. All these
representations are well known and classified.
Given a complex representation $\varphi$ of $G_c$ that does not admit a real
form for $G$, one may form a real representation for $G$ in
$\varphi+\overline{\varphi}$, with $\overline{\varphi}$ the complex conjugate
representation of $\varphi$, called the {\em realification} of $\varphi$.
\item
If the complex representation $\varphi$ of $G_c$ admits a real form for $G$
then the ring of $\varphi(G)$-invariant polynomials, with real coefficients,
$\real [ \real^n]^{\varphi (G)}$, admits a real IB.
The ring of $\varphi(G_c)$-invariant polynomials, with complex coefficients,
is exactly the ring obtained from $\real [ \real^n]^{\varphi (G)}$
allowing the coefficients of the polynomials to be complex numbers:
$$ \complex [ \complex^n ]^{\varphi (G_c)}
\simeq  \complex  \otimes \real [ \real^n ]^{\varphi (G)}$$
(The space $\real^n$ or
$\complex^n$ where the polynomials are defined is irrelevant here,
as one may consider that all of them are defined in terms of $n$
abstract indeterminates).\\
Then, both $ \complex [ \complex^n ]^{\varphi (G_c)}$ and
$\real [ \real^n ]^{\varphi (G)}$ admit the same IB formed by
polynomials with real coefficients.
\item
A representation $\varphi(G)$ in $\real^n$ is coregular if and only if the
representation $\varphi(G_c)$ in $\complex^n$ is coregular. In fact
any algebraic relation $\hat f$ among the $p_i(x)$ (it doesn't matter if the
coefficients in $\hat f$ are real or complex) implies that both $\varphi(G)$
and $\varphi(G_c)$ are non-coregular.
\item
Every subrepresentation $\varphi'$ of a coregular representation $\varphi$,
is coregular too.
The basic invariants of $\varphi'$ are a proper subset of the basic
invariants of $\varphi$.
\end{itemize}
From the classification of G. W. Schwarz \cite{sch} then one
may recover the classification of the real coregular representations of the
compact simple Lie groups. All what one has to do is to select in
\cite{sch} the representations of the complex simple Lie
groups $G_c$ that are complexifications
of real representations of the maximal compact subgroups $G$ of $G_c$.
This gives the classification of all coregular real linear representations
of compact simple Lie groups, together with the number and degrees of
the basic invariants, and it is reported in \cite{stv2}.
(The IB of the real linear groups $\varphi(G)$ and of their complexifications
$\varphi(G_c)$ can be chosen to be real and the same for the two groups).

The next step is to determine the orbit spaces, or equivalently the
$\widehat P$-matrices, of the real coregular representations of compact simple
Lie groups.\\
For each real coregular representation $\varphi$ of a compact simple
Lie group that have an IB with $q\leq 4$ basic invariants,
we select all possible candidates in the list of the allowable
$\widehat P$-matrices given in table 1 that have the same
number and degrees of the basic invariants.
In some cases we are
left with only one possibility, but in some others there are more different
choiches.\\
When there are more than one candidate $\widehat P$-matrix,
we must select among the candidates, the right one.
In the case  of adjoint representations the choice is easy, because
in this case the orbit space is the same of that of the corresponding
Weyl group \cite{stv} and one already knows the $\widehat P$-matrices
of the irreducible finite reflection groups. In all other
cases the hints given by Schwarz to construct the IB
are sufficient to determine the right choice, even if the IB is not known
completely. These calculations are reported in \cite{stv}.

Table~\ref{tabFG} lists the $\widehat P$-matrices of the irreducible
representations of coregular finite groups with $q\leq 4$ basic invariants
and tables~\ref{tabLG1} and \ref{tabLG2} list the $\widehat P$-matrices of coregular
representations of compact simple Lie groups with $q\leq 4$ basic invariants.

To denote the irreducible representation with maximal weight
$\Lambda=(\Lambda_1,\Lambda_2,\ldots)$ I shall use the notation:
${\varphi_1}^{\Lambda_1}{\varphi_2}^{\Lambda_2}\ldots$, omitting to write
${\varphi_i}^{\Lambda_i}$ when ${\Lambda_i}=0$ and omitting to write
${\Lambda_i}$ when ${\Lambda_i}=1$.
The notation here differs sometimes with that reported in several texts
on group theory for the ordering of the roots in the Dynkin diagrams but
I prefere here to maintain the same notation of \cite{sch}.
As an aid to the reader in the tables \ref{tabLG1} and \ref{tabLG2}
it is reported also the dimension of the representation,
so no ambiguity can occour.\\
In the following tables~\ref{tabLG1} and \ref{tabLG2}
in the column $G$ the (compact, real) Lie group is indicated by
the symbol of the Lie algebra of its complexification,
and this means that it is written $A_n$ for $SU(n+1)$,
$B_n$ for $SO(2n+1)$, $C_n$ for $Sp(2n)$, $D_n$ for $SO(2n)$.\\
In the cases of one or two invariants there is only one class of
allowable $P$-matrices (beside equivalences).
I shall denote with $I_1$ and $I_2$
the classes of $\widehat P$-matrices of dimension 1 and 2 respectively.\\
In tables~\ref{tabLG1} and \ref{tabLG2}, to avoid isomorphisms, the ranks are limited in the
following way: $B_n,n\ge 2$; $C_n,n\ge 3$; $D_n,n\ge 4 $.
Representations that differ only for changes:
of $\varphi_i \leftrightarrow \varphi_{n-i}$, $i=1,\ldots,[\frac{n}{2}]\ $
for $A_n$,
for changes: $\varphi_{n-1} \leftrightarrow  \varphi_n\ $ for $D_n$,
for permutations of $\varphi_1$, $\varphi_3$, $\varphi_4\ $ for $D_4$,
are avoided.

\section{Conclusions}
The main conclusions of our calculations are as follows:
\begin{enumerate}
\item
It is possible to classify all allowable $\widehat P$-matrices of compact
coregular linear groups \cite{6st,7st} or of compact non-coregular linear
groups of $r$-type $(q,q-1)$ \cite{stv}. These matrices determine univocally
the sets ${\cal S}\subset \real^q$ that represent the orbit spaces.
\item
This classification is done without using the integrity basis and without
knowing any specific information of group structure, but using
only some very general algebraic conditions.
\item
All existing compact linear groups determine $\widehat P$-matrices of
the same form (eventually after an integrity basis transformation)
of an allowable $\widehat P$-matrix.
When for a given set of the degrees $d_1,\ldots,d_q$ there are no allowable
$\widehat P$-matrix, then there are also no compact linear group with the basic
invariants of those degrees.
\item
Finite groups and compact Lie groups may share the same orbit space structure.
\end{enumerate}
The main open problems in all this subject are the following:
\begin{enumerate}
\item
Given an allowable $\widehat P$-matrix $\widehat P$, does it always exist a compact linear
group whose integrity basis defines $\widehat P$? When this group exists,
which is the group and its integrity basis.
\item
What is the meaning of the induction rules and what is their relation with
group theory?
\item
Is it always true that the allowable $\widehat P$-matrices of non-coregular groups
of $r$-type $(q,q-1)$, that is with only one relation among the basic
invariants, coincide with the allowable $\widehat P$-matrices of the coregular
groups?
\end{enumerate}

The results here reviewed are partial but they point out
a very strong relation with group theory and with
invariant theory which ought to be further investigated.\\
One fact that appears clearly from the table~\ref{tabR}
is that different representations of different groups may share the same
orbit space. The orbit spaces of finite groups and of Lie groups may also
be the same. This happens because the orbit spaces are determined only by the
$\widehat P$-matrices, that is only by the way how the scalar products between the
gradients of the basic invariants $p_i(x)$ are expressed in terms of the
$p_i(x)$. When for two integrity basis these expressions are the same,
then the $\widehat P$-matrices and the orbit spaces are the same.
From table~\ref{tabR} one sees that this happens often, even
considering only coregular groups.

Some future work might be oriented towards the following goals:
\begin{enumerate}
\item
Find the 5-dimensional allowable $\widehat P$-matrices of coregular groups.
This will clarify the induction rules.
\item
Find the 4- and 5-dimensional allowable $\widehat P$-matrices of non-coregular
groups of $r$-type $(q,q-1)$. This will clarify the link between the
coregular and non-coregular case.
\item
Find the groups generating the allowable $\widehat P$-matrices.
\item
Study if and how the link between a group and one of its subgroups or the
link between a direct product group and its factor groups gives some
links also between the corresponding $\widehat P$-matrices.
\end{enumerate}

\begin{table}[h]\begin{center}           
\caption{ORBIT SPACES OF REAL COREGULAR REPRESENTATIONS OF COMPACT
SIMPLE LIE GROUPS WITH $q=1, 2, 3, 4$ BASIC INVARIANTS. I.} \label{tabLG1}
\vskip 0.4cm\begin{tabular}{|c|c|c|c|c|c|c|c|c|}
\hline
Entry
   &$G$      &    $\varphi$         &{\it dim}&i/r& $q$& $d_i$   &\#$P$  &$P$  \\
\hline
1  &$A_1$    &  $2 \varphi_1    $   &    4    &  i  & 1 &  2      & 1 &  $I_1$            \\
2  &         & $ {\varphi_1}^2=Ad$  &    3    &  i  & 1 &  2      & 1 &  $I_1$            \\
3  &         & $ 2{\varphi_1}^2$    &  3+3    &  r  & 3 &  2,2,2  & 2 &  $I(1,1)$     \\
4  &         & $ {\varphi_1}^4$     &    5    &  i  & 2 &  3,2    & 1 &  $I_2$            \\
5  &$A_{n\ge2}$
         & $\varphi_1+\varphi_n$
                    &$2(n+1)$ &  i  & 1 &  2      & 1 &  $I_1$            \\
6  &         & $ 2\cdot(\varphi_1+\varphi_n)$
                    &$2\cdot2(n+1)$
                          &  r  & 4 & 2,2,2,2 & 5 & $A1(1,1,1,1)$ \\
7  &$A_2$    & $ \varphi_1 \varphi_2=Ad$
                    & 8       &  i  & 2 & 3,2     & 1 & $I_2$             \\
8  &         & $ \varphi_1^2+ \varphi_2^2$
                    & 12      &  i  & 4 & 4,3,3,2 & 1 &  $B2$        \\
9  &$A_3$    & $ \varphi_1 \varphi_3=Ad$
                    & 15      &  i  & 3 & 4,3,2   & 1 &  $III.1$     \\
10 &         & $ \varphi_2$         & 6       &  i  & 1 &  2      & 1 & $I_1$        \\
11 &         & $ \varphi_1+ \varphi_2+ \varphi_3$
                    & 8+6     &  r  & 2 & 2,2     & 1 & $I_2$        \\
12 &         & $ 2\varphi_2$        & 6+6     &  r  & 3 & 2,2,2   & 2 & $I(1,1)$     \\
13 &$A_4$    & $ \varphi_1 \varphi_4=Ad$
                    & 24      &  i  & 4 & 5,4,3,2 & 1 &  $E1$        \\
14 &         & $ \varphi_2+\varphi_3$
                    & 20      &  i  & 2 & 4,2     & 1 & $I_2$     \\
15 &$A_5$    & $ \varphi_2+\varphi_4$
                    & 30      &  i  & 4 & 4,3,3,2 & 1 & $B2$     \\
16 &$A_6$    & $ \varphi_2+\varphi_5$
                    & 42      &  i  & 3 & 6,4,2   & 3 & $III.2$     \\
17 &$A_8$    & $ \varphi_2+\varphi_7$
                    & 72      &  i  & 4 & 8,6,4,2 & 4 & $E3$     \\
18 &$B_{n\ge2}$
         & $ \varphi_1$         & $2n+1$  &  i  & 1 & 2       & 1 & $I_1$\\
19 &         & $ 2\varphi_1$        &$2\cdot(2n+1)$
                          &  r  & 3 & 2,2,2   & 2 & $I(1,1)$\\
20 &$B_2$    & $ \varphi_1^2$       & 14      &  i  & 4 & 5,4,3,2 & 1 & $E1$     \\
21 &         & $ \varphi_2=Ad$      & 10      &  i  & 2 & 4,2     & 1 & $I_2$\\
22 &$B_3$    & $ \varphi_2=Ad$      & 21      &  i  & 3 & 6,4,2   & 3 & $III.2$\\
23 &         & $ \varphi_3$         &  8      &  i  & 1 & 2       & 1 & $I_1$\\
24 &         & $\varphi_1+\varphi_3$&  7+8    &  r  & 2 & 2,2     & 1 & $I_2$\\
25 &         & $2\varphi_1+\varphi_3$
                    &  7+7+8  &  r  & 4 & 2,2,2,2 & 5 & $A3(1,1,1)$\\
26 &         & $ 2\varphi_3$        &  8+8    &  r  & 3 & 2,2,2   & 2 & $I(1,1)$\\
27 &$B_4$    & $ \varphi_2=Ad$      &  36     &  i  & 4 & 8,6,4,2 & 4 & $E3$\\
28 &         & $ \varphi_4$         &  16     &  i  & 1 & 2       & 1 & $I_1$\\
29 &         & $ \varphi_1+\varphi_4$
                    &  9+16   &  r  & 3 & 3,2,2   & 2 & $I(2,1)$\\
30 &         & $ 2\varphi_4$        &  16+16  &  r  & 4 & 4,2,2,2 & 10& $A1(1,1,2,2)$\\
31 &$D_{n\ge4}$
         & $ \varphi_1$         & $2n$    &  i  & 1 & 2       & 1 & $I_1$\\
32 &         & $ 2\varphi_1$        &$2n+2n$  &  r  & 3 & 2,2,2   & 2 & $I(1,1)$\\
33 &$D_4$    & $ \varphi_2=Ad$      & 28      &  i  & 4 & 6,4,4,2 & 7 & $E2$\\
34 &         & $ \varphi_1+\varphi_3$
                    &  8+8    &  r  & 2 & 2,2     & 1 & $I_2$\\
35 &         & $ \varphi_1+\varphi_3+\varphi_4$
                    &  8+8+8  &  r  & 4 & 3,2,2,2 & 6 & $A2(1,1,1,1)$\\
36 &         & $ \varphi_1+2\varphi_3$
                    &  8+8+8  &  r  & 4 & 2,2,2,2 & 5 & $ A3(1,1,1)$\\
37 &$D_5$    & $ \varphi_4+\varphi_5$
                    &  32     &  i  & 2 & 4,2     & 1 & $I_2$\\
\hline\end{tabular}\end{center}
\end{table}

\begin{table}[h]\begin{center}           
\caption{ORBIT SPACES OF REAL COREGULAR REPRESENTATIONS OF COMPACT
SIMPLE LIE GROUPS WITH $q=1, 2, 3, 4$ BASIC INVARIANTS. II.} \label{tabLG2}
\vskip 0.4cm\begin{tabular}{|c|c|c|c|c|c|c|c|c|}
\hline
Entry
   & $\ G\ $ &    $\qquad \varphi\qquad $ &$\ $\quad{\it dim}\quad$\ $&i/r& $q$& $d_i$   &\#$P$    &$\qquad P \qquad $\\
\hline

38 &$C_3$    & $ 2\varphi_1    $    &  12     &  i  & 1 & 2       & 1 & $I_1$\\
39 &         & $ \varphi_1^2=Ad$    &  21     &  i  & 3 & 6,4,2   & 3 & $III.2$\\
40 &         & $ \varphi_2     $    &  14     &  i  & 2 & 3,2     & 1 & $I_2$\\
41 &$C_4$    & $ \varphi_1^2=Ad$    &  36     &  i  & 4 & 8,6,4,2 & 4 & $E3$\\
42 &         & $ \varphi_2     $    &  27     &  i  & 3 & 4,3,2   & 1 & $III.1$\\
43 &$C_5$    & $ \varphi_2     $    &  44     &  i  & 4 & 5,4,3,2 & 1 & $E1$\\
44 &$E_6$    & $ \varphi_1+\varphi_5$
                    &  54     &  i  & 4 & 4,3,3,2 & 1 & $B2$\\
45 &$F_4$    & $ \varphi_1$         &  26     &  i  & 2 & 3,2     & 1 & $I_2$\\
46 &         & $ \varphi_4=Ad$      &  52     &  i  & 4 & 12,8,6,2& 3 & $E4$\\
47 &$G_2$    & $ \varphi_1$         &  7      &  i  & 1 & 2       & 1 & $I_1$\\
48 &         & $ 2\varphi_1$        &  7+7    &  r  & 3 & 2,2,2   & 2 & $I(1,1)$\\
49 &         & $ \varphi_2=Ad$      &  14     &  i  & 2 & 6,2     & 1 & $I_2$\\
\hline\end{tabular}\end{center}
{\bf Notation for tables \ref{tabLG1}, \ref{tabLG2}}.
Entry: line number.
$G$: Compact Lie group (indicated by the Lie algebra of its complexification).
$\varphi$: real representation of $G$.
{\it dim}: real dimension of $\varphi$.
i/r: reducibility.
$q$: number of basic invariants.
$d_i$: degrees of the basic invariants.
\#$P$: number of different allowable $\widehat P$-matrices with degrees $d_i$.
$P$: $\widehat P$-matrix and corresponding orbit space of the linear group
$(G,\varphi)$.
\end{table}

\begin{table}[h]\begin{center}           
\caption{ORBIT SPACES OF REAL IRREDUCIBLE REPRESENTATIONS OF COREGULAR
FINITE GROUPS WITH $q=1, 2, 3, 4$ BASIC INVARIANTS. } \label{tabFG}
\vskip 0.4cm\begin{tabular}{|c|c|c|c|c|}
\hline

$G$      & $dim=q$& $d_i$   &\#$P$  &$P$\\
\hline
$A_1$    &     1 &  2      & 1 &  $I_1$            \\
$I_2(m)$ &     2 & $m$,2     & 1 & $I_2$             \\
$A_3$    &     3 & 4,3,2   & 1 &  $III.1$     \\
$B_3$    &     3 & 6,4,2   & 3 & $III.2$\\
$H_3$    &     3 & 10,6,2  & 3 & $III.3$\\
$A_4$    &     4 & 5,4,3,2 & 1 &  $E1$        \\
$D_4$    &     4 & 6,4,4,2 & 7 & $E2$\\
$B_4$    &     4 & 8,6,4,2 & 4 & $E3$\\
$F_4$    &     4 & 12,8,6,2& 3 & $E4$\\
$H_4$    &     4 & 30,20,12,2 & 2 & $E5$\\
\hline\end{tabular}\end{center}
{\bf Notation}.
$G$: Finite group. {\it dim}: dimension of the representation.
$q$: number of basic invariants.
$d_i$: degrees of the basic invariants.
\#$P$: number of different allowable $\widehat P$-matrices with degrees $d_i$.
$P$: $\widehat P$-matrix and corresponding orbit space of the group $G$.
\end{table}

\begin{table}
\begin{center} \label{tabR}
\caption{ORBIT SPACES OCCOURING IN TABLES 2, 3 AND 4}
\vskip 0.4cm\begin{tabular}{|c|c|cc|}
\hline
 $P$-matrix     &$d_i$   &      &    $G$            \\
\hline
$I_1          $&  2     &          $A_1$,& $<\!1\!>$, $<\!2\!>$, $<\!5\!>$, $<\!10\!>$, $<\!18\!>$, $<\!23\!>$,\\
           &        &          & $<\!28\!>$, $<\!31\!>$, $<\!38\!>$, $<\!47\!>$\\
$I_2          $&  $d$,2 &$I_2(d)$,& $<\!4\!>$, $<\!7\!>$, $<\!11\!>$, $<\!14\!>$, $<\!21\!>$, $<\!24\!>$, \\
           &        &          & $<\!34\!>$, $<\!37\!>$, $<\!40\!>$, $<\!45\!>$, $<\!49\!>$\\
$I(1,1)       $&2,2,2   &          & $<\!3\!>$, $<\!12\!>$, $<\!19\!>$, $<\!26\!>$, $<\!32\!>$, $<\!48\!>$\\
$I(2,1)       $&3,2,2   &          & $<\!29\!>$\\
$III.1        $&4,3,2   &$A_3$,    &  $<\!9\!>$, $<\!42\!>$\\
$III.2        $&6,4,2   & $B_3$,    & $<\!16\!>$, $<\!22\!>$, $<\!39\!>$\\
$III.3        $&10,6,2  & $H_3$    & \\
$A1(1,1,1,1)  $&2,2,2,2 &           &$<\!6\!>$\\
$A1(1,1,2,2)  $&4,2,2,2 &           &$<\!30\!>$\\
$A2(1,1,1,1)  $&3,2,2,2 &           &$<\!35\!>$\\
$A3(1,1,1)    $&2,2,2,2 &           &$<\!25\!>$, $<\!36\!>$\\
$B2           $&4,3,3,2 &           &$<\!8\!>$, $<\!15\!>$, $<\!44\!>$\\
$E1           $&5,4,3,2 &$A_4$,     & $<\!13\!>$, $<\!20\!>$, $<\!43\!>$\\
$E2           $&6,4,4,2 &$D_4$,     & $<\!33\!>$\\
$E3           $&8,6,4,2 &$B_4$,     & $<\!17\!>$, $<\!27\!>$, $<\!41\!>$\\
$E4           $&12,8,6,2&$F_4$,     & $<\!46\!>$\\
$E5           $&30,20,12,2& $H_4$   &\\
\hline\end{tabular}\end{center}
{\bf Notation}.
$P$-matrix: Type of $\widehat P$-matrix.
$d_i$: degrees of the basic invariants.
$G$: linear group, indicated by its symbol if it is a finite group or by its
entry number in tables~\ref{tabLG1} and \ref{tabLG2} if it is a Lie group.
\end{table}

\end{document}